\documentclass[preprint2]{aastex}

\shorttitle{A Supercluster at $z\sim$0.91 in SSA22}
\shortauthors{Kim et al.}

\begin{document}


\title{Discovery of a Supercluster at $z\sim$0.91 and Testing the $\Lambda$CDM Cosmological Model}


\author{Jae-Woo Kim\altaffilmark{1,2}, Myungshin Im\altaffilmark{1,2}, Seong-Kook Lee\altaffilmark{1,2}, 
Alastair C. Edge\altaffilmark{3}, Minhee Hyun\altaffilmark{1,2}, Dohyeong Kim\altaffilmark{1,2}, 
Changsu Choi\altaffilmark{1,2}, Jueun Hong\altaffilmark{1,2}, Yiseul Jeon\altaffilmark{2}, Hyunsung David Jun\altaffilmark{2,4}, 
Marios Karouzos\altaffilmark{2}, Duho Kim\altaffilmark{2,5}, Ji Hoon Kim\altaffilmark{6}, Yongjung Kim\altaffilmark{1,2}, 
Won-Kee Park\altaffilmark{7}, Yoon Chan Taak\altaffilmark{1,2}, and Yongmin Yoon\altaffilmark{1,2}}
\email{kjw0704@gmail.com, mim@astro.snu.ac.kr}


\altaffiltext{1}{Center for the Exploration of the Origin of the Universe, Department of Physics and Astronomy, Seoul 
National University, Seoul 151-742, Republic of Korea}
\altaffiltext{2}{Astronomy Program, FPRD, Department of Physics and Astronomy, Seoul National University, Seoul 151-742, Republic of Korea}
\altaffiltext{3}{Department of Physics, University of Durham, South Road, Durham DH1 3LE, UK}
\altaffiltext{4}{Jet Propulsion Laboratory, California Institute of Technology, 4800 Oak Grove Drive, Pasadena, CA 91109, USA}
\altaffiltext{5}{Arizona State University, School of Earth and Space Exploration, P.O. Box 871404, Tempe, AZ 85287-1404, USA}
\altaffiltext{6}{Subaru Telescope, National Astronomical Observatory of Japan, 650 North A'ohoku Place, Hilo, HI 96720, USA}
\altaffiltext{7}{Korea Astronomy and Space Science Institute, Daejeon 305-348, Republic of Korea}


\begin{abstract}

The $\Lambda$CDM cosmological model successfully reproduces many aspects of the galaxy and structure formation of the universe. 
However, the growth of large-scale structures (LSSs) in the early universe is not well tested yet with observational data. 
Here, we have utilized wide and deep optical--near-infrared data in order to search for distant galaxy clusters and superclusters ($0.8<z<1.2$).  
From the spectroscopic observation with the Inamori 
Magellan Areal Camera and Spectrograph (IMACS) on the Magellan telescope, three massive clusters at $z\sim$0.91 are confirmed in the SSA22 field. 
Interestingly, all of them have similar redshifts within $\Delta z\sim$0.01 with velocity dispersions ranging from 470 to 1300 km s$^{-1}$.
Moreover, as the maximum separation is $\sim$15 Mpc, they compose a supercluster at $z\sim$0.91, meaning that  
this is one of the most massive superclusters at this redshift to date. The galaxy density map implies that the confirmed clusters are 
embedded in a larger structure stretching over $\sim$100 Mpc. 
$\Lambda$CDM models predict about one supercluster like this in our surveyed volume, consistent with our finding so far. 
However, there are more supercluster candidates in this field, suggesting that 
additional studies are required to determine if the $\Lambda$CDM cosmological model can successfully 
reproduce the LSSs at high redshift.

\end{abstract}





\section{INTRODUCTION}

Under the currently popular $\Lambda$CDM cosmology \citep{im97, rie98, per99, eis05}, the large-scale structure (LSS) of galaxies emerges when 
the initial density fluctuations grow with time through gravitational attraction between galaxies. The $\Lambda$CDM cosmological model has been successful in reproducing 
the LSS at $z\simeq0$, showing the promise of the $\Lambda$CDM cosmology to explain our universe \citep[e.g.,][]{bah03,wil11,ben13}.

However, the growth of LSSs has not been thoroughly tested yet with observational data at $z\gg0$. LSSs stretch from several tens of Mpc to a few hundred Mpc, 
but there is a lack of data sets that are deep and wide enough to cover such structures at high redshifts. So far, high-redshift LSS studies have 
been limited mostly to galaxy-cluster-scale structures ($\lesssim$1--2 Mpc), with mixed results. Some results show agreements with the $\Lambda$CDM cosmology 
models \citep{wil11, bay14}, but others suggest too many massive clusters at $z\gtrsim1$ \citep{jee09, gon12, kan15}.

With the advance of large and deep imaging surveys, it is now possible to extend the test of the cosmological formation of LSSs to scales much larger  
than before. Specific predictions have been made about superclusters at high redshift. A supercluster represents the most massive structure in the 
universe with sizes of up to $\sim$100--200 Mpc, containing filaments, multiple galaxy clusters and groups. 
Several studies have pointed out that superclusters are useful objects to test cosmological models \citep{wra06,ein11,lim14}.

So far, only a handful number of superclusters have been found at $z\sim1$.
The Cl 1604 supercluster at $z\sim$0.91 has 8 member clusters and groups that have velocity dispersions in the range of  
$\sim$280 to 820 km s$^{-1}$ \citep{lub00,gal08,lem12,asc14,wu14}. Another supercluster is identified at $z\sim0.89$ in the Elais-N1 field, containing 5 
clusters \citep{swi07}. A compact supercluster, RCS 2319+00 \citep{gil08}, stands as
the most massive supercluster found at $z\sim$0.9 with the summed mass of members exceeding $10^{15}\,M_{\odot}$ \citep{fal13}, and 
a separation between the member clusters is less than 3 Mpc. Finally, the Lynx supercluster at $z\sim1.26$ contains two X-ray clusters 
and three groups confirmed spectroscopically \citep{ros99,mei12}.
 
In order to unveil LSSs and other interesting high-redshift objects, we have been conducting the Infrared Medium-deep Survey (IMS; Im, M., et al. 
2016, in preparation). The IMS is a deep ($J\sim23$ AB mag) and wide ($\sim 120$ deg$^{2}$) near-infrared (NIR) imaging survey that combines deep $J$-band 
imaging data with other optical/NIR survey data, making it possible to find LSSs at $z\sim1$. Here, we report the discovery of a new, massive 
supercluster at $z\sim0.9$ in the SSA22 area as the first result, and discuss if such an LSS is compatible with cosmological simulation predictions.

We adopt cosmological parameters for the flat universe with $\Omega_{m}=$0.27, $H_{0}$=71 km s$^{-1}$ Mpc$^{-1}$, and $\sigma_{8}=$0.8. All magnitudes 
are in the AB system. In addition, all distance scales are physical scales based on the angular diameter distance, unless otherwise noted.

\section{DATA AND CLUSTER FINDING}\label{datclu}

\subsection{{\it Photometric Catalog}}\label{photcat}

Our work is based on wide and deep data sets for the SSA22 field 
($\alpha=$22$^{h}$17$^{m}$00$^{s}$ and $\delta=$00\arcdeg20\arcmin00\arcsec) from the Canada-France-Hawaii Telescope (CFHT) Legacy Survey 
(CFHTLS\footnote{http://www.cfht.hawaii.edu/Science/CFHTLS/}), the United Kingdom Infrared Telescope (UKIRT) Infrared Deep Sky Survey \citep[UKIDSS;][]{law07} 
Deep eXtragalactic Survey (DXS; Edge et al. 2016, in preparation), and the IMS. Although UKIDSS 
DXS ($J$- and $K$-bands) and IMS mapped nearly the entire CFHTLS--W4 ($ugriz$-bands) area (25 deg$^2$), the effective area is $\sim$20 deg$^2$ after excluding 
regions such as halos and spikes of bright stars. 
The 80\% point-source detection limits are $u^{*}\sim25.2$, $g'\sim25.6$, $r'\sim25.0$, $i'\sim24.9$, and $z'\sim23.9$ for the CFHTLS\footnote{http://terapix.iap.fr/}, 
and $J\sim23.7$ and $K\sim23.2$ for UKIDSS DXS and IMS \citep{kim11, kim15}.
Sources were detected using SExtractor \citep{ber96} in dual mode, and 
the unconvolved $J$-band images were used for the detection and to measure the $J$-band total magnitude. In addition, 2\arcsec\,diameter 
apertures were applied to PSF matched images to derive aperture magnitudes for the color measurement. For this work, we applied a magnitude cut of 
$J=23.2$ that is the 90\% point-source completeness limit of the $J$-band data. Photometric redshifts ($z_{phot}$) were 
derived using the {\it Le Phare} software \citep{arn99,ilb06} after training the data using spectroscopic redshifts with flags of 3, 4, 23, and 24 
from the VIMOS VLT Deep Survey \citep[VVDS;][]{lef05,gar08}. A measured redshift accuracy ($\sigma_{pz}$) by the normalized median absolute deviation is 
$\Delta z/(1+z)=0.038$, and the outlier fraction is $<$5\%. 
Details of the procedures are described in \citet{kim15}.

\subsection{{\it Finding MSGs}}\label{finclu}

Using all objects in the photometric redshift catalog, we searched for massive structures of galaxies (MSGs, galaxy clusters and groups) 
between $z=0.8$ and $z=1.2$.
After splitting galaxies into redshift bins from $z_{bin}=0.8$ to $z_{bin}=1.2$ with an increment of 0.02 and 
a bin size of $|z_{phot}-z_{bin}|<\sigma_{pz}(1+z_{bin})$ based on the best-fit photometric redshift, 
the Voronoi Tessellation technique \citep{ebe93,soa11} is applied to measure a local density ($\rho=1/$area$_{\rm cell}$) for each galaxy, which is 
converted into the normalized cell density, $\delta$ ($=\rho/\rho_{\rm median}$). 
Galaxies are identified to be in an overdense region if its $\delta$ value is above a threshold, $\delta_{\rm thres}$, following the prescription 
described in \citet{soa11}. The $\delta_{\rm thres}$ values are determined for each redshift bin, and they are found to vary between 
$\delta=1.9$ and $2.0$. The threshold corresponds to approximately 4$\sigma$ above the mean density, if we fit the $\delta$ distribution with a 
Gaussian function in linear scale at $\delta<1$\footnote{This constraint is adopted to avoid the contribution of LSSs.}. 
We group galaxies whose cells have $\delta>\delta_{\rm thres}$ and are adjacent to each other as a possible overdense area.
Then, we classify the overdense area as an MSG candidate if it has a probability over 95\% that the signal is not due to random fluctuations 
\citep[Eq. (3) of][]{soa11}. The sky position of a MSG candidate is assigned as the coordinate of the galaxy with the highest density. 
Also, the redshift is taken to be the median redshift of its galaxies within 1 Mpc from the MSG candidate center.
MSG candidates from different redshift bins are merged into a single candidate if the projected separation between them is less than 2 Mpc, and 
their redshift bins overlap with each other. 

We estimated the fraction of bona fide member galaxies in this approach using a galaxy mock catalog of {\tt GALFORM} \citep{col00,mer13}. 
For this, we randomly scatter mock galaxy redshifts with $\sigma_{pz}$ above, and then select galaxies 
with $|z_{phot}-z_{cen}|<\sigma_{pz}(1+z_{cen})$ and 
within 1.0 and 1.5 Mpc radii from central galaxies of 346 massive halos at $0.85<z_{cen}<1.15$, where $z_{cen}$ indicates the halo redshift. The average 
fractions of bona fide members among selected galaxies are 50\% and 30\% for 1.0 and 1.5 Mpc, respectively.
Therefore, we select MSG candidates, only if they have at least 
25 galaxies ($N_{\rm 1.5 Mpc}\ge25$) within a 1.5 Mpc radius and the photometric redshift uncertainty and with $J<J^{*}+1$ (where $J^{*}$ is 
characteristic magnitude). In total, there are 691 MSG candidates.

In order to identify supercluster candidates, we count the number of MSG candidates within a 10 Mpc radius and the photometric redshift 
uncertainty from each MSG candidate. Then, supercluster candidates are chosen as a group of at least 10 MSG candidates. 
Through this process, we find two supercluster candidates at the median photometric redshift of $\sim$0.89 and another at $\sim$0.92.

\section{CLUSTER CONFIRMATION}

\subsection{{\it IMACS Observation and Redshift Determination}}\label{speobs}

Multi-object spectroscopy was performed on 2014 September 23, using the Inamori Magellan Areal Camera and Spectrograph (IMACS) on 
the Magellan/Baade telescope in its f/2 mode of a field that covers a 27\arcmin.4 diameter field of view at 
$\alpha=$22$^{h}$13$^{m}$08$^{s}$ and $\delta=$00\arcdeg40\arcmin24\arcsec. Of supercluster candidates described in the previous section, 
we chose the target field due to its unusually high 
concentration in a small area: six prominent ($N_{\rm 1.5 Mpc}>40$) and nine less significant ($N_{\rm 1.5 Mpc}<40$) MSG candidates, within 
a photometric redshift range of $0.85<z_{phot}<0.96$. 
Slitlets were assigned to galaxies in prominent candidates first, and then to those in less significant candidates.
In order to choose target galaxies for the spectroscopy, 
we used the probability distribution function (PDF) of photometric redshifts from the {\it Le Phare} software. The integrals 
of the normalized PDFs within the uncertainty range ($\sigma_{pz}$) from the candidate redshift were calculated as the 
probability for each galaxy belonging to the cluster \citep{bru00, pap10, bro13}.
Galaxies with probabilities $>$0.5 were selected as potential members. Spectra of potential members were taken using the 200 lines mm$^{-1}$ 
grism with the WB5600--9200 filter ranging from 5600\AA~to 9200\AA. One slit mask was used for the observation with 1\arcsec$\times$6\arcsec\,slitlets. In total, 320 slitlets were assigned 
for galaxies including potential {\b MSG members} (80\%) and field galaxies (20\%). The spectral resolution was $\lambda/\Delta\lambda\sim$600.
The total on-source integration time was 2.5 hours (30 minutes$\times$5) under $\sim$0\arcsec.9 seeing.

We used the Carnegie Observatories System for MultiObject Spectroscopy (COSMOS\footnote{http://code.obs.carnegiescience.edu/cosmos}) to reduce the 
IMACS spectroscopic data. The procedure includes standard reduction algorithm, wavelength calibration, and sky subtraction. We extracted one-dimensional 
spectra for each source from two-dimensional spectra stacked by the COSMOS pipeline. The flux calibration was performed using an F5-type star that 
was also included in the slit mask. 

The redshift of each galaxy was determined with the {\it SpecPro} software \citep{mas11}. We mainly used the emission and 
absorption lines of [OII]3727, Ca H\&K, the 4000\AA\,break, the G-band, and the Balmer lines (H${\delta}$ and H${\gamma}$) for this. 
If only a single emission line was detected, we considered this as the [OII] line. If the identified line was not [OII], e.g., such as 
H$\gamma$, H$\beta$, [OIII] or H$\alpha$, it would likely be accompanied by another line at shorter or longer wavelengths and be at a 
redshift that is difficult to explain the continuum shape (i.e., at a redshift that is very different from photometric redshifts). 
We successfully determined the 
redshift of 217 galaxies, implying a success rate of $\sim$70\%. The success rate is $\sim$80\% for galaxies at $i_{AB}\leq22.5$.
Among successful spectroscopic measurements, 51\% comes from a single emission, and 7\% comes from Ca H\&K absorption lines.

\subsection{{\it Discovery of Supercluster at $z=0.91$}}\label{clupro}

Using the galaxies with spectroscopic redshifts, we determine the membership of each galaxy. For this process, we follow 
the iterative algorithm described in \citet{lub02}.
First, we select galaxies within a 1 Mpc radius from the cluster position determined by the photometric redshift method in \S~\ref{finclu}. Then, we calculate 
the bi-weight mean ($z_{cl}$) and scale ($\sigma_{z}$) of redshifts of these galaxies \citep{bee90}. We exclude galaxies with 
$\left | z-z_{cl} \right | >3\sigma_{z}$ or the relative rest-frame radial velocity greater than 3500 km\,s$^{-1}$. This process is repeated until 
no more galaxies are excluded. Finally, the dispersion ($\sigma'_{z}$) is calculated by the gapper method due to the small number of 
members and then converted into the velocity dispersion ($\sigma_{v}$). 
Uncertainties in $\sigma_{v}$ are estimated by the Jackknife resampling.
Through this process, we identify three galaxy clusters at $z\sim0.91$ with $\sigma_{v}>$470 km\,s$^{-1}$, which corresponds to 
$M_{200}>1.1\times10^{14} M_{\odot}$ at the cluster redshift \citep{car97, dem10}. Table~\ref{tcl} lists the properties of the confirmed clusters 
based on two different radii of 1 Mpc and 1.5 Mpc for the comparison.
Independently, we estimated the uncertainty in $\sigma_{v}$ by randomly selecting 100 times 11 and 7 spectroscopic members within a 
1 Mpc radius of the centers of two galaxy clusters at $z\sim1.2$ containing $\sim$20-30 galaxies each with $\sigma_{v}=$490-650 km s$^{-1}$ 
\citep{muz09}. We find that the standard deviation from this exercise to be 160 km s$^{-1}$ and 320 km s$^{-1}$ for the 11 and 7 member cases, 
respectively, which are consistent with or smaller than the Jackknife resampling errors in Table~\ref{tcl}.

Figure 1a shows example spectra of cluster members in the three confirmed clusters. Vertical lines indicate locations of 
[OII] (blue), Ca H\&K (red), and H$\delta$ (magenta) lines. 
Figure 1b displays the spectroscopic redshift distribution of confirmed members for each cluster. 
The fraction of members of which spectroscopic redshifts are determined by a single line is 55\%, 29\%, and 0\% for 
IMSCl J2212+0045, IMSCl J2213+0052 and IMSCl J2213+0048, respectively. 
Note that the velocity dispersion based only on red galaxies could decrease as much as  $\sim$50\% of that from both blue and red galaxies \citep{gal08}. 
When we did a similar analysis on two clusters where at least 5 and 4 red galaxies ($(r-i)>0.9$ and $(i-z)>0.5$) are available 
(J2212+0045 and J2213+0048), $\sigma_{v}$ becomes 200$\pm$149 and 1117$\pm$464 km s$^{-1}$, respectively. This suggests that the $\sigma_{v}$ values in 
Table~\ref{tcl} could be overestimated.
Figure~\ref{figimg} shows pseudo color images for the confirmed clusters with photometric (circles) and spectroscopic (squares) members.

For the SSA22 field, \citet{dur11} also identified candidate galaxy clusters using photometric redshifts derived from the optical CFHTLS data. 
Of our clusters, IMSCl J2213+0052 and IMSCl J2213+0048 have counterparts in their list, considering a matching radius of 
3\arcmin~($\sim$1.5 Mpc at $z=0.91$) and $\Delta z<0.1$.

Interestingly, all the new clusters are massive and located at redshifts of $z\sim$0.908--0.920. In addition, the maximum projected angular 
separation between galaxy clusters is 16\arcmin.78, which corresponds to $\sim$15 Mpc (comoving) at the cluster redshift. The proximity of the three massive clusters 
suggests that this is a supercluster. 
Compared to galaxy clusters in the Cl 1604 supercluster with a maximum velocity dispersion of $\sim$800 km s$^{-1}$, 
the supercluster presented here may encompass one or more clusters that are more massive and is 
more comparable to the RCS 2319+00 supercluster members (710--1200 km s$^{-1}$) and members in the Elais-N1 field (660--1000 km s$^{-1}$).
Additionally, the maximum projected angular separation is similar to that for 
the three most massive clusters in Cl 1604 (16\arcmin.24), larger than RCS 2319+00 (7\arcmin.62), and smaller than the Elais-N1 supercluster (32\arcmin.45). 
This new supercluster in the SSA22 field may be one of the most massive structures ever found at $z\sim$0.9.
We summarize the properties of the known superclusters in Table~\ref{tcl}.

\section{DISCUSSION}

\subsection{{\it Large-scale Structure}}\label{lss}

In order to see if there are more structures beyond the confirmed clusters through a galaxy density map, we apply the Voronoi tessellation 
technique described in \S~\ref{finclu} to the redshift bin at $z=0.914$, the mean redshift of the confirmed clusters.
Then, we make a grid map with a grid cell size of 500 kpc across the entire SSA22 area and calculate the mean value of local densities 
($\delta$ in \S~\ref{finclu}) for each grid.

Figure~\ref{figcont} shows the map of local densities at $z\sim$0.914 around the confirmed clusters over a 3.5 deg $\times$ 2.9 deg area 
($\sim$190 Mpc $\times$ 160 Mpc, comoving). 
Small and large red circles indicate the positions of the newly confirmed clusters and the IMACS pointing, respectively. 
Intriguingly, the density contours extend toward the south and northeast with 
the structure spanning from (R.A., decl.)$=$(333.12, -0.1) to (334.15, 1.5). 
The projected angular size of this structure is $\sim$114\arcmin.2 corresponding to $\sim$54 Mpc or $\sim$103 Mpc (comoving) at this 
redshift. We also overlay galaxies (orange points) with 
spectroscopic redshifts of 0.90$<z_{spec}<$0.92 from the VVDS and the VIMOS Public Extragalactic Redshift Survey 
that are deemed reliable \citep[flags 2--9 and 22--29; VIPERS;][]{gar14,guz14}.
Comparing the density map and the distribution of spectroscopic samples, it seems that the supercluster extends the northeast direction.
Note that no spectroscopic redshifts are available in the southwest area. 
The density map suggests that the supercluster possibly extends to a much larger scale.

\subsection{{\it Comparison with Models}}\label{mod}

Here, we examine if the existence of the new supercluster at $z=0.91$ can be explained with $\Lambda$CDM models.

First, we search for dark matter halos grouped similarly to our confirmed clusters from the Millennium simulation \citep{spr05} 
with the WMAP-7 cosmology \citep{guo13}. We use 15 snapshots from 
$z=1.77$ to $z=0.51$, each with a 0.32 Gpc$^{3}$ cube (i.e., five times the volume of our data at $0.8<z<1.2$). 
At $z=1.08$, the first structures with properties comparable to the new supercluster form, i.e., those containing at least 
three halos, each in excess of $M_{200}>1.1\times10^{14}M_{\odot}$. Two such structures appear at this redshift. 
By $z=0.8$, three or four such structures have formed.
When translated to our survey volume, the simulation suggests $\sim$0.6--0.8 superclusters at 
$0.8<z<1.2$. If we use the Millennium simulation with $\sigma_{8}=0.9$, the number is comparable in this redshift range. 

Second, we calculate the predicted number of superclusters based on the supercluster mass function of \citet{lim14}. 
As a conservative estimate, we set the supercluster mass at $10^{15}M_{\odot}$. Using their mass function at $z=1$, the predicted number of 
superclusters with $>10^{15}M_{\odot}$ in the 20 deg$^{2}$ area with $0.8<z<1.2$ is $\sim$0.6. 

The expected numbers of superclusters from the models are consistent with the number of superclusters we identified so far. However, 
there are still two more supercluster candidates in the SSA22 field, and extended structures as discussed in \S~\ref{lss} may contain more 
superclusters. On the other hand, considering the uncertainty of cluster masses, the clusters can be lighter. If so, the observation and the model 
predictions can be reconciled since lighter superclusters are more abundant than heavier ones in models. 
To understand if there is any tension between observed superclusters and $\Lambda$CDM models, 
it is necessary to do a more thorough analysis of 
larger cosmological simulations and an intensive spectroscopic mapping of these large structures.

\acknowledgments

Authors thank an anonymous referee for comments that were useful for improving the paper.
This work was supported by the National Research Foundation of Korea (NRF) grant, No. 2008-0060544, funded
by the Korea government (MSIP).
ACE acknowledges support from STFC grant ST/L00075X/1.
D.K. acknowledges fellowship support from the grant NRF-2015-Fostering Core Leaders of Future Program, No. 2015-000714, funded by the Korean government.
M.H. acknowledges the support from Global PH.D Fellowship Program through the National Research Foundation of Korea (NRF) funded by the Ministry of Education 
(NRF-2013H1A2A1033110). 
We are grateful to UKIDSS team, the staff in UKIRT, Cambridge Astronomical Survey Unit and Wide Field
Astronomy Unit in Edinburgh. The United Kingdom Infrared Telescope was run by the Joint Astronomy Centre on
behalf of the Science and Technology Facilities Council of the U.K. 
This work is based in part on data products produced at the Canadian Astronomy Data Centre as part
of the Canada-France-Hawaii Telescope Legacy Survey, a collaborative project of NRC and CNRS.
This Letter uses data from the VIMOS Public Extragalactic Redshift Survey (VIPERS). VIPERS has been performed 
using the ESO Very Large Telescope, under the "Large Programme" 182.A-0886. The participating institutions 
and funding agencies are listed at http://vipers.inaf.it.
This research uses data from the VIMOS VLT Deep Survey, obtained from the VVDS database operated by Cesam, 
Laboratoire d'Astrophysique de Marseille, France.
Finally, authors also thank staff in the Las Campanas Observatory.



{\it Facilities:} \facility{Magellan (IMACS)}, \facility{UKIRT (WFCAM)}, \facility{CFHT}.

\clearpage

\begin{table*}
\begin{center}
\caption{Summary of confirmed clusters in this study with applying two different radii (top) and previously reported superclusters at $z\sim0.9$ (bottom).\label{tcl}}
\resizebox{\linewidth}{!}{
\begin{tabular}{ccccccccc}
\tableline\tableline
Cluster&R.A. (J2000)&decl. (J2000)&Radius (Mpc)&$n_{\rm slit}$&$n_{\rm member}$&$z_{cl}$&$\sigma_{v}$ (km s$^{-1}$)&$M_{200}$ ($\times10^{14}M_{\odot}$)\\
\tableline
IMSCl J2212+0045&22:12:28&00:45:06&1.0&19&11&0.9170$\pm$0.0008&474$\pm$152&1.1$_{-0.8}^{+1.5}$\\
&&&1.5&33&13&0.9171$\pm$0.0022&584$\pm$148&2.1$_{-1.2}^{+2.0}$\\
IMSCl J2213+0052&22:13:02&00:52:02&1.0&13&7&0.9196$\pm$0.0026&884$\pm$469&7.2$_{-6.5}^{+18.7}$\\
&&&1.5&26&9&0.9176$\pm$0.0028&944$\pm$305&8.8$_{-6.1}^{+11.6}$\\
IMSCl J2213+0048&22:13:31&00:48:42&1.0&17&7&0.9085$\pm$0.0041&1298$\pm$310&23.0$_{-12.8}^{+20.7}$\\
&&&1.5&25&11&0.9118$\pm$0.0036&1665$\pm$329&48.5$_{-23.4}^{+34.8}$\\
\tableline
\end{tabular}
}
\resizebox{\linewidth}{!}{
\begin{tabular}{ccccccccc}
\tableline\tableline
Supercluster&R.A.$^{a}$&decl.$^{a}$&$z$&$n_{\rm cluster}$&$\sigma_{v}$$^{b}$&Size$_{\rm projected}$$^{b}$&$<n_{\rm galaxy}>$$^{b}$&Ref.\\
\tableline
Cl 1604&16:04:23&43:13:08&0.85--0.94&8&688--818&17\arcmin&$\sim$70&Wu+14$^{c}$\\
&&&&&590--811&&$\sim$51&Gal+08$^{c}$\\
Swinbank+07&16:08:27&54:35:47&0.89&5&730--1030&32\arcmin&$\sim$11&Swinbank+07\\
RCS 2319+00&23:19:53&00:38:04&0.90&8&714--1202&8\arcmin&$\sim$16&Faloon+13\\
\tableline
\end{tabular}
}
\tablenotetext{a}{Coordinate for the most massive cluster.}
\tablenotetext{b}{Values based on the three most massive clusters.}
\tablenotetext{c}{Member galaxies within two times the virial radius in \citet{wu14} and 1 $h^{-1}$Mpc in \citet{gal08}.}
\end{center}
\end{table*}



\begin{figure}
\epsscale{1.0}
\plotone{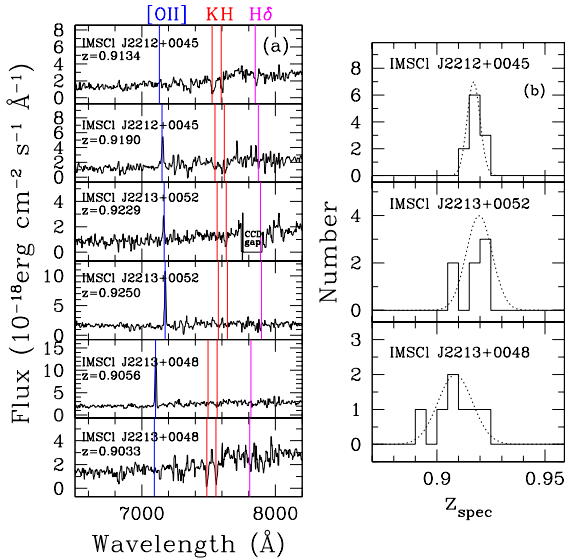}
\caption{{\bf (a)} Example IMACS spectra of the confirmed cluster members. The cluster IDs and the spectroscopic redshifts are noted in each panel. 
The vertical lines mark the [OII]3727, Ca H\&K and H${\delta}$4102 lines at the noted redshift. 
{\bf (b)} Redshift distribution of spectroscopic members within 1 Mpc. The dotted lines show Gaussian distributions 
based on $z_{cl}$ and $\sigma'_{z}$ (see \S~\ref{clupro} for details).
\label{figspzh}}
\end{figure}

\begin{figure}
\epsscale{0.7}
\plotone{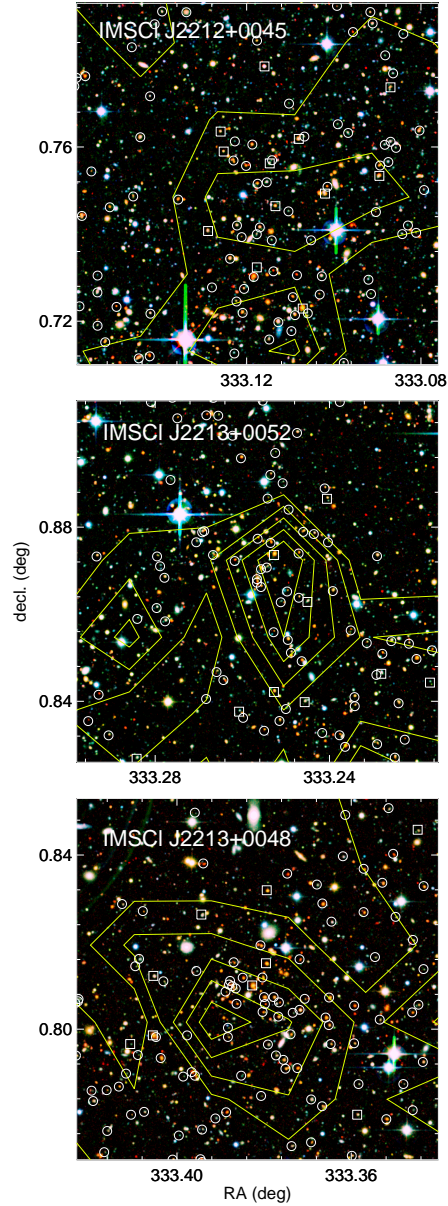}
\caption{Pseudo-color images ($giK$-bands) of the confirmed clusters. These images were arbitrarily scaled for display purposes. 
The field of view is 5\arcmin$\times$5\arcmin\, corresponding to 2.4 Mpc$\times$2.4 Mpc 
at the cluster redshift. The circles and the squares are for photometric and spectroscopic members, respectively. 
Yellow curves show contours for $<\delta>=$2, 4, 6, 7, and 10 at $z\sim0.91$ (\S~\ref{lss}).\label{figimg}}
\end{figure}

\begin{figure*}
\epsscale{1.6}
\plotone{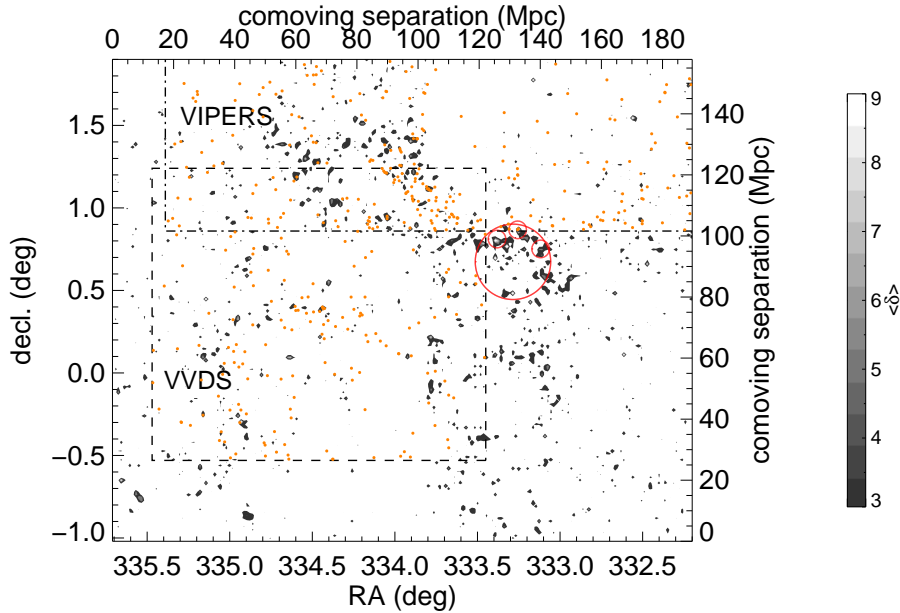}
\caption{ Overdensity contour at $z=0.914$ around the confirmed clusters. 
Small and large red circles show the confirmed clusters and the IMACS field of view, respectively. Dashed and dotted-dashed boxes indicate survey 
boundaries of VVDS and VIPERS, respectively. 
The orange points are galaxies with spectroscopic redshifts between $z=$0.90 and 0.92 from VVDS and VIPERS. 
More MSG candidates exist at northeast of the supercluster, and spectroscopic samples from VVDS and VIPERS seem to connect the supercluster 
and the MSG candidates. 
\label{figcont}}
\end{figure*}

\clearpage

\end{document}